\documentclass[preprint,,authoryear,12pt]{elsarticle}
\usepackage{amssymb}
\usepackage{graphicx}
\usepackage{srcltx}

\journal{New Astronomy Reviews}
\begin{document}
\begin{frontmatter}

\title{Rydberg atoms in astrophysics}

\author[cao]{Yu.~N.~Gnedin}
\author[if]{A.~A.~Mihajlov}
\ead{mihajlov@ipb.ac.rs}
\author[if]{Lj.~M.~Ignjatovi{\' c}}
\author[if]{N.~M.~Sakan}
\author[if]{V.~A.~Sre{\' c}kovi{\' c}}
\author[StP]{M.~Yu.~Zakharov}
\author[StP]{N.~N.~Bezuglov}
\author[StP]{A.~N.~Klycharev}

\address[cao]{Central astronomical observatory, Pulkovo, RAS}
\address[if]{Institute of physics, P.O. Box 57, 11001 Belgrade, Serbia}
\address[StP]{Department of Physics, Saint-Petersburg University, Ulianovskaya 1, 198504 St. Petersburg, Petrodvorets, Russia}

\begin{abstract}
Elementary processes in astrophysical phenomena traditionally
attract researchers attention. At first this can be attributed to a
group of hemi-ionization processes in Rydberg atom collisions with
ground state parent atoms. This processes might be studied as a
prototype of the elementary process of the radiation energy
transformation into electrical one. The studies of nonlinear
mechanics have shown that so called regime of dynamic chaos should
be considered as typical, rather than exceptional situation in
Rydberg atoms collision. From comparison of theory with experimental
results it follows that a such kind of stochastic dynamic processes,
occurred during the single collision, may be observed.
\end{abstract}


\end{frontmatter}
\section{Introduction}

If an atom is in a state of sufficiently high principal quantum
number $n$ it means that the valence electron is far from the ionic
core, and such atom appears hydrogenic. In such atoms the valence
electron is influenced mainly by the positive charge of the ionic
core, and not by its structure. The excited state of these hydrogen
like atoms are commonly accepted to call Rydberg states, high
Rydberg states, or simply highly excited states \citep{prvi}. For
example the radius of Bohr orbit of Rydberg atom in the state
$n=110$ is equal to $6.4 \cdot 10^{-5}\ \textrm{cm}$, i.e.~very
close to macroscopic size. Another obvious property of such atom is
the large transition dipole moment, which simply reflects the
separation between the valence electron and the positive ionic core.
Such atoms, as a result, provide the extremely high cross sections
of the relevant physical processes.

Although the study of Rydberg atoms has a long history, the
development of laser technique has led to a great experimental
advances and in recent years renewed interest for such researches.
Another very important field of researches for Rydberg atoms is
astrophysics. The basic goal of our paper is description of a
various astrophysical situations that are responsible for production
of Rydberg states and their observational effects. Our paper
consists of three parts: (a) short review of astrophysical target of
Rydberg atoms (b) atom- Rydberg atom chemi-ionization collisional
processes in stellar atmospheres, and (c) description of dynamic
chaos regime in Rydberg atoms collisions.

\subsection{The base astrophysical targets of searches for Rydberg
atoms}

Some models of big bang nucleosynthesis suggest that very high
baryon density regions were formed in the early universe. Then the
most important aspect is hydrogen recombination and Rydberg states
can play important role in this process. It is well known
\citep{drugi} that the chemical composition of the primordial gas
consists of electrons and:

\begin{itemize}
\item hydrogen: $H$, $H^{-}$, $H^{+}$, $H_{2}^{+}$ and $H_{2}$
\item deuterium: $D$, $D^{+}$, $HD$, $HD^{+}$ and $HD^{-}$
\item helium: $He$, $He^{+}$, $He^{2+}$ and $HeH^{+}$
\item lithium: $Li$, $Li^{+}$, $Li^{-}$, $LiH^{-}$ and $LiH^{+}$
\end{itemize}

\begin{figure}
\includegraphics[angle=0, width=\textwidth]{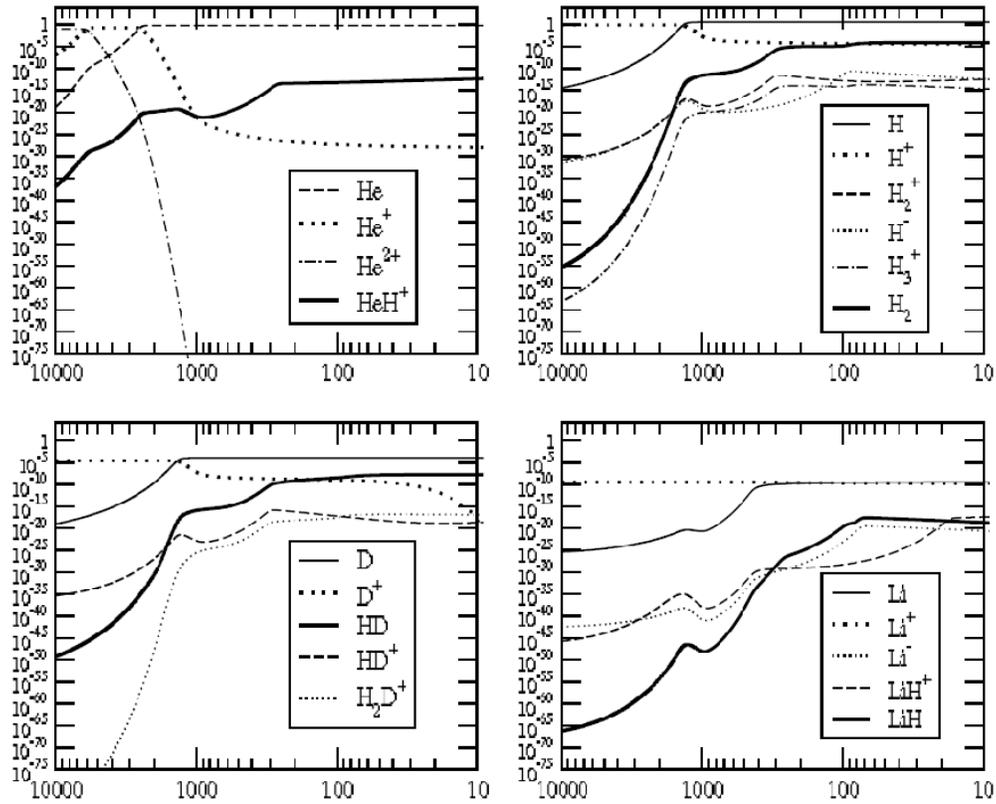}
\caption{Evaluation of chemical abundances in the standard Big Bang model.
Vertical axes are the relative abundances and the horizontal axes are
relative to the redshift}
\label{fig::1}
\end{figure}

Their respective abundances are calculated from a set of chemical
reactions for the early universe \citep{drugi}. Evaluation of
chemical abundances in the standard BB model is presented at Fig.
\ref{fig::1} from \citep{drugi}.

\begin{figure}
\includegraphics[angle=0, width=\textwidth]{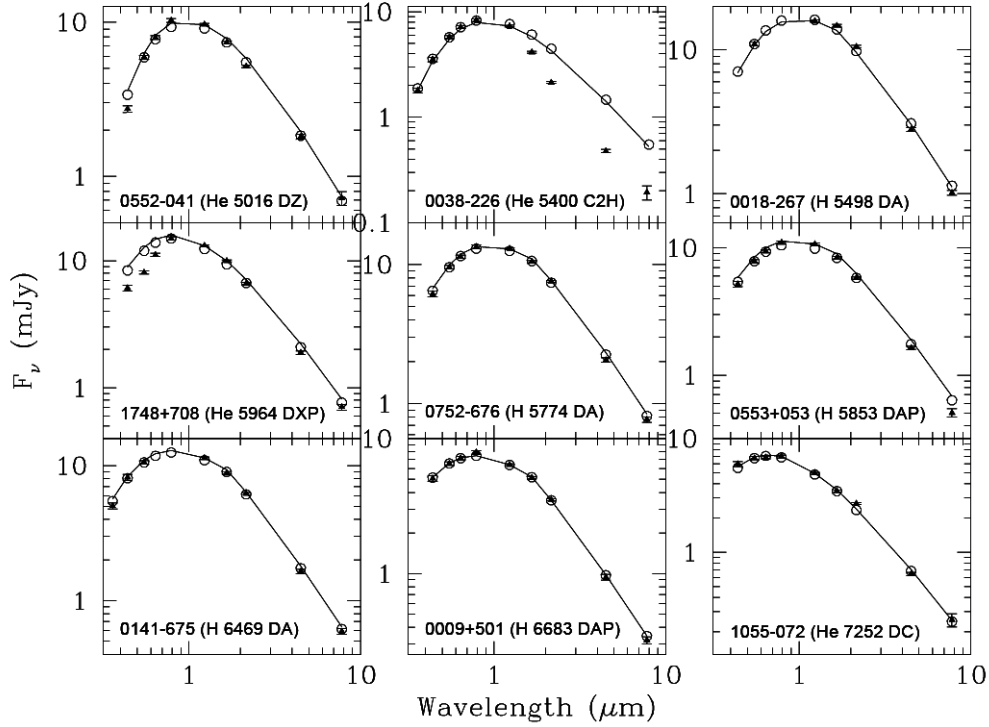}
\caption{Spectral energy distributions of cool white dwarfs observed
with Spitzer space telescope ($1 \textrm{Jy} = 10^{-26}\cdot
\textrm{W} \cdot \textrm{m}^{-2} \cdot \textrm{Hz}^{-1}$).}
\label{fig::2}
\end{figure}

Another probable regions of Rydberg atoms existence are the cool
stars and, firstly, the cool white dwarfs. Recently, a new effect
has been noticed from Spitzer observations of cool white dwarfs
\citep{treci}. Namely, these observations have demonstrated that
some white dwarfs with $T < 6100\ K$ are found to display
significant flux deficits in Spitzer observations, (see Figure
\ref{fig::2} from \citet{treci}). These mid-IR flux deficits are not
predicted by the current white dwarf models including collision
induced absorption due to molecular hydrogen. This fact implies that
the source of this flux deficit is not standard molecular absorption
but some other physical process. We claim that such process may be
absorption by atoms and molecules in highly excited Rydberg states.

The similar effect have been recently discovered and for magnetic
white dwarfs from spectropolarimetric observations made at Russian
BTA-6m telescope and NIR photometric observations of magnetic white
dwarfs at Russian-Italian AZT-94 telescope located at Campo
Imperatore. The details can be found in \citet{cetvrti}.

\section{Atom – Rydberg atom chemi-ionization collision processes in stellar
atmospheres}
Besides of the star atmospheres with strong magnetic fields, the
processes with the participation of the Rydberg's atoms could be of
interest in the atmospheres of other types of stars where the
presence of magnetic field can be neglected in the first
approximation. As an example, within this study is considered the
significance of the chemi-ionization processes

\begin{equation}\label{eq:jedan}
\begin{array}{c}
   H^{*}(n) + H \rightarrow e + H^{+} + H \\%
   H^{*}(n) + H \rightarrow e + H^{+}_{2}
\end{array}
\end{equation}

\begin{equation}\label{eq:dva}
\begin{array}{c}
   He^{*}(n) + He \rightarrow e + He^{+} + He \\%
   He^{*}(n) + He \rightarrow e + He^{+}_{2}
\end{array}
\end{equation}

\noindent and inverse chemi-recombination processes

\begin{equation}\label{eq:tri}
\begin{array}{c}
   e + H^{+} + H \rightarrow H^{*}(n) + H   \\%
   e + H^{+}_{2} \rightarrow H^{*}(n) + H
\end{array}
\end{equation}

\begin{equation}\label{eq:cetiri}
\begin{array}{c}
   e + He^{+} + He \rightarrow  He^{*}(n) + He  \\%
   e + He^{+}_{2} \rightarrow He^{*}(n) + He
\end{array}
\end{equation}

\noindent for some stelar atmospheres. Here $H$  and $He$ are the
atoms in the ground states, $H^{*}(n)$ and $He^{*}(n)$ - the Rydberg
atoms (in the states with the principal quantum number $n \gg 1$),
$H_{2}^{+}$ and $He_{2}^{+}$ - the molecular ions in the ground
electronic states, and e - free electron.

In accordance with \citet{peti} and \citet{sesti} it is assumed
that:

-   the processes (\ref{eq:jedan},\ref{eq:dva}) occur at the parts
of the trajectories of the atom-projectile, $H$ or $He$, which lie
deeply inside the orbit of the outer electron in the Rydberg atom,
$H^{*}(n)$ or $He^{*}(n)$ and are caused by dipole part of the
interaction of this electron with the ion-atom complex $H+H^{+}$ or
$He+He^{+}$;

-  the considered processes are treated as the result of the
resonant conversion of the energy within the electron component of
the collision atom-Rydberg atom system, what understands that the
transition of the outer electron from the initial bound (Rydberg)
state with given $n$ to the final free state with some momentum $k$
occurs simultaneously with the transition of ion-atom complex from
the first excited electronic state to the ground electronic state;

- the processes can be described by means of the decay velocity of
the initial electronic states which depends only from the
internuclear distance.

From the results of \citet{peti} and \citet{sesti} it follows that
the processes (\ref{eq:jedan}-\ref{eq:cetiri}) are significant for
such hydrogen and helium plasmas with $\frac{Ne}{Nat} < 10^{-3}$,
where $Ne$ and $Nat$ are the free electron and ground state atom
density. This means that these processes could be significant for
the stellar atmospheres which contain the corresponding weakly
ionized layers. Here, this assumption is illustrated by the results
related to one of M red star (in the hydrogen case) and to some of
DB white dwarfs (in the helium case).

\subsection{M red star}

In \citet{sedmi} it is shown that the processes (\ref{eq:jedan}) and
(\ref{eq:tri}) in the region of $n$ from 4 to 8 influence to the
populations of all hydrogen atom excited states with $n>3$ what is
illustrated by next figures \ref{fig:prvas} and \ref{fig:drugas},
which show the behavior of the ratio of the excited states
population calculated with and without these processes.

\begin{figure}
  \includegraphics[width=\textwidth, height=0.8\textwidth]{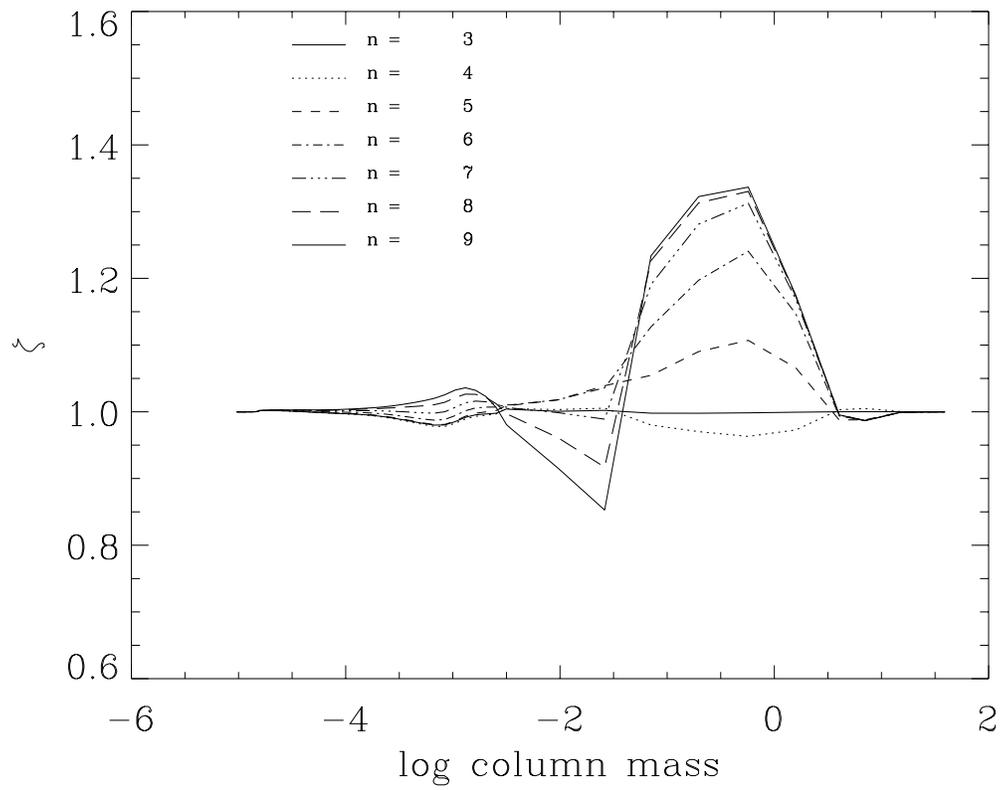}\\
  \caption{The behavior of the population ratio $\zeta$  for $3\leq n \leq 9$
           as a function of the column mass.}
   \label{fig:prvas}
\end{figure}

\begin{figure}
  \includegraphics[width=\textwidth, height=0.8\textwidth]{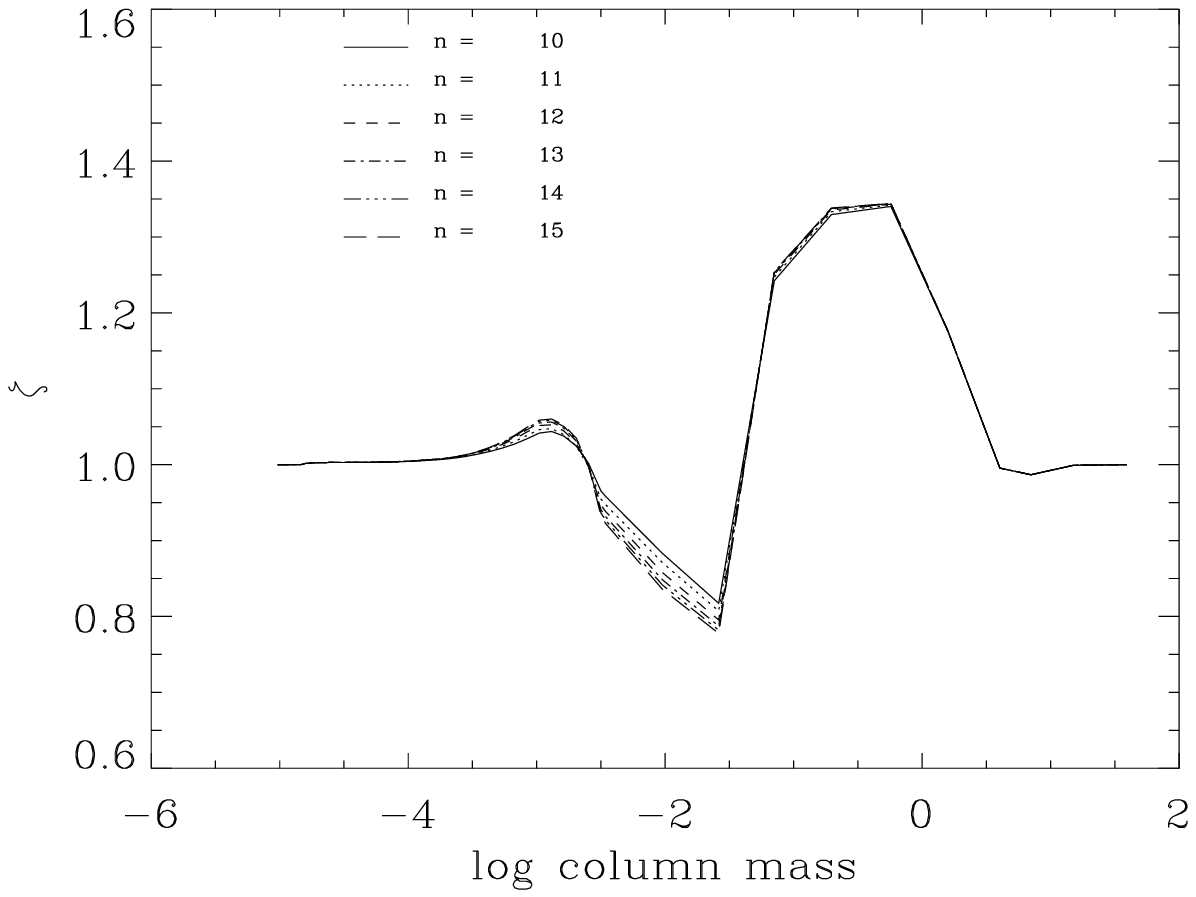}\\
  \caption{The behavior of the population ratio $\zeta$  for $3\leq n \leq 9$
           as a function of the column mass.}
   \label{fig:drugas}
\end{figure}

\begin{figure}
  \includegraphics[width=\textwidth, height=0.8\textwidth]{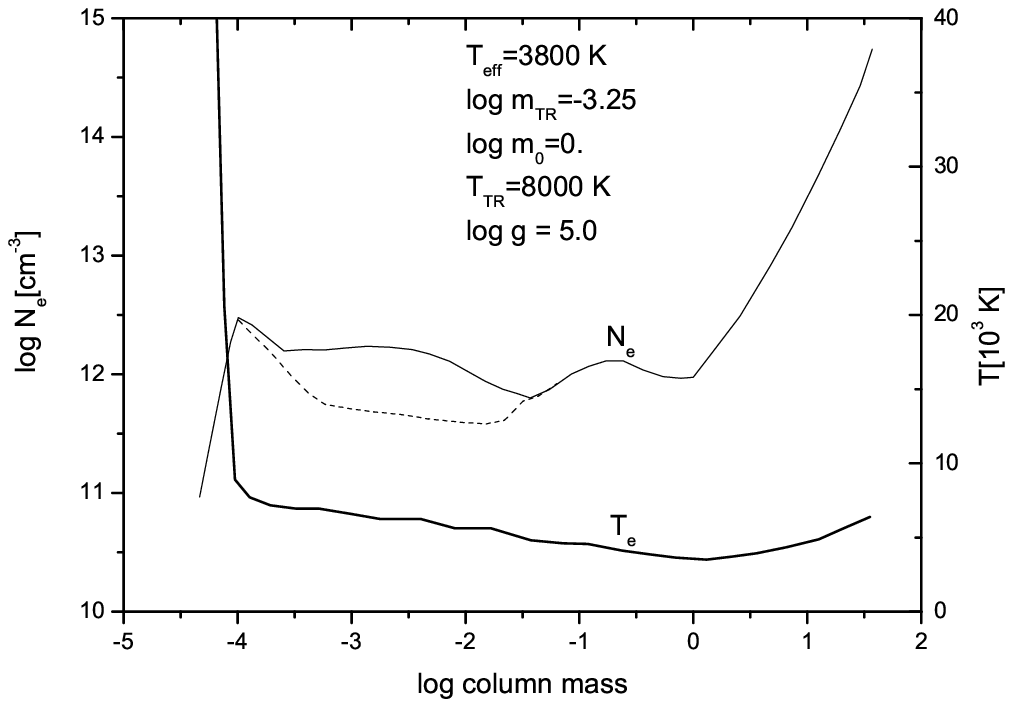}\\
  \caption{Structure of model atmosphere - temperature $T_{e}$  and electron density
           $N_{e}$  vs. column mass.}
   \label{fig:trecas}
\end{figure}

Then, in \citet{deveti} is shown that the processes (\ref{eq:jedan})
and (\ref{eq:tri}) in the whole region of $n>1$ also influence to
the free electron density what is illustrated by figure
\ref{fig:trecas}, which shows the behavior of free electron density
calculated with these processes (solid curve) and without of them
(dashed curve) .

\begin{figure}
  \includegraphics[width=\textwidth, height=0.8\textwidth]{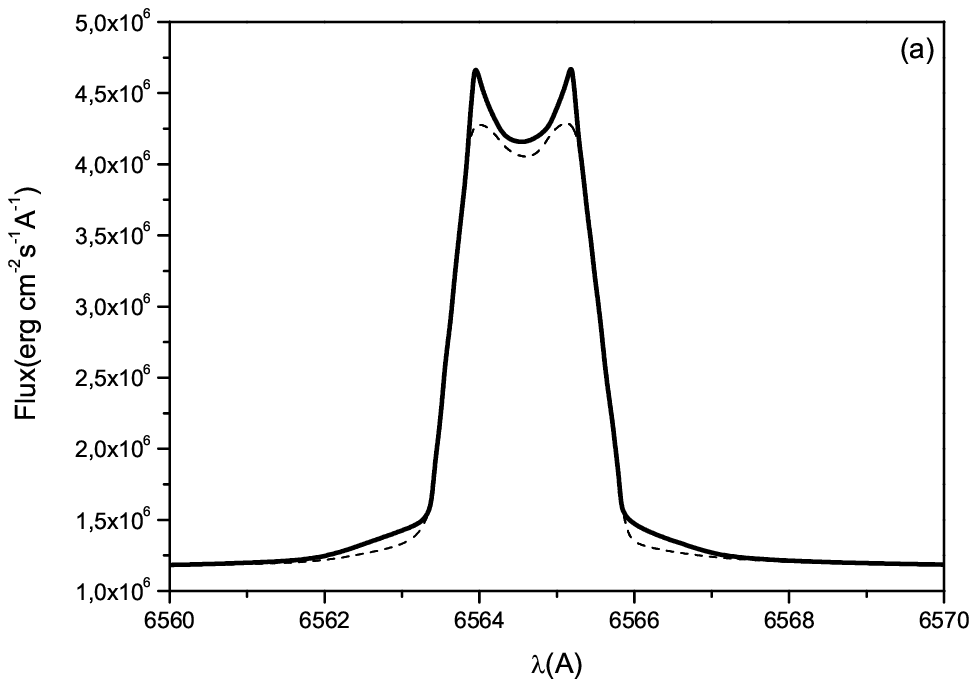}\\
  \caption{Line profiles with (full) and without (dashed) inclusion of chemi-ionization and
  chemi-recombination processes for $H_{\alpha}$  line}
   \label{fig:cetvrtas}
\end{figure}

\begin{figure}
  \includegraphics[width=\textwidth, height=0.8\textwidth]{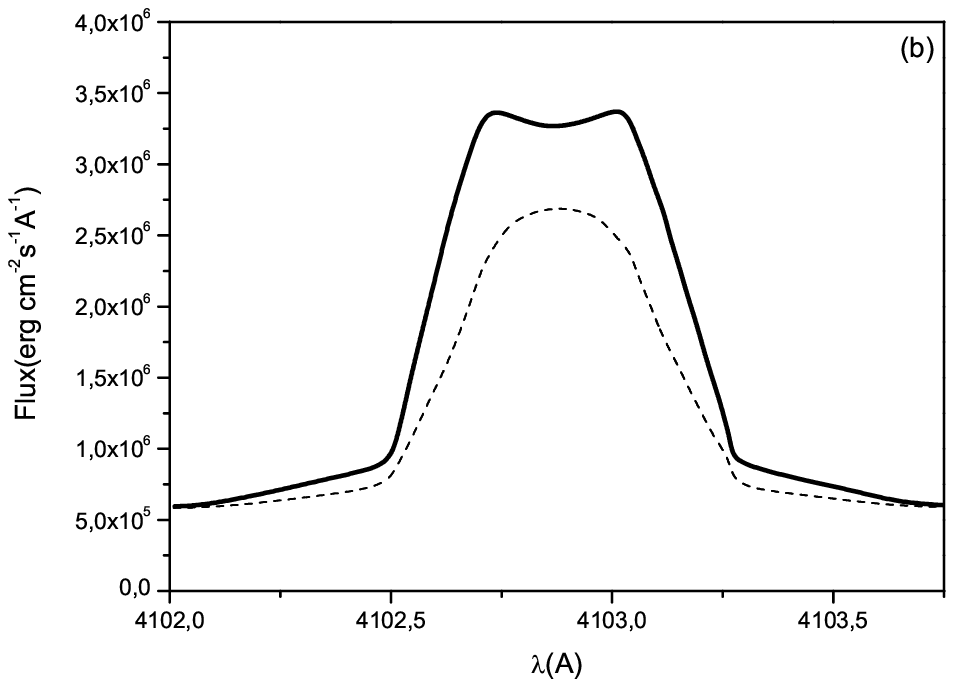}\\
  \caption{Line profiles with (full) and without (dashed) inclusion of chemi-ionization
  and chemi-recombination processes for $H_{\beta}$   line}
   \label{fig:petas}
\end{figure}

The presented results suggest that the processes (\ref{eq:jedan})
and (\ref{eq:tri}), due to their influence on the excited state
populations and the free electron density, also should influence on
the atomic spectral line shapes. This assumption is confirmed by the
figures \ref{fig:cetvrtas} and \ref{fig:petas}, which show the
profiles of some of hydrogen spectral lines calculated with and
without these processes.

The presented results suggest that the processes (\ref{eq:jedan})
and (\ref{eq:tri}), due to their influence on the excited state
populations and the free electron density, also should influence on
the atomic spectral line shapes. This assumption is confirmed by the
figures, which show the profiles of some of hydrogen spectral lines
calculated with and without these processes.

\subsection{DB white dwarfs}

\begin{figure}
  \includegraphics[width=\textwidth, height=0.8\textwidth]{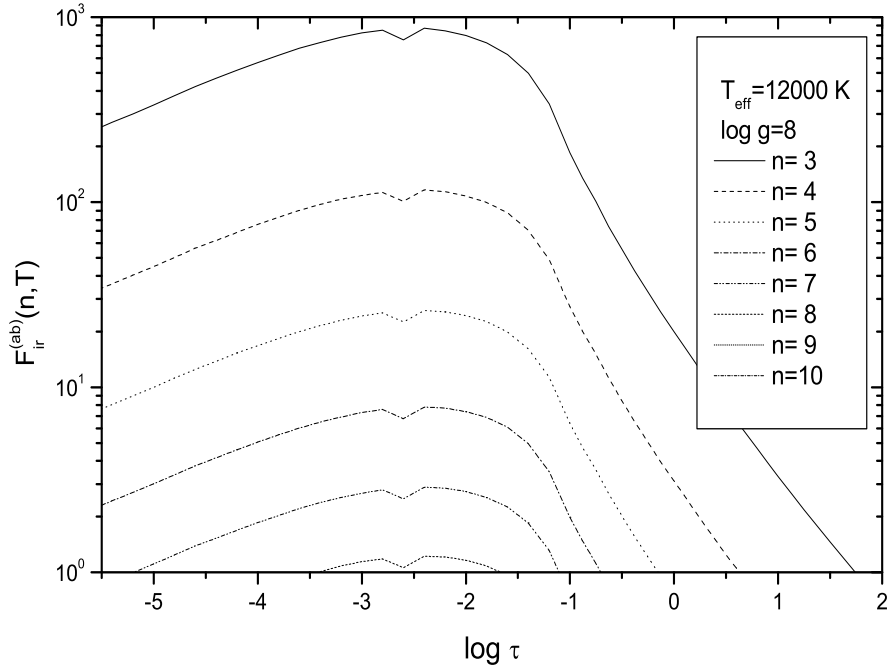}\\
  \caption{Parameter $F_{ir}^{(ab)}(n,T)$ as a function of the logarithm of Rosseland optical depth
  $\textrm{log}(\tau)$, for principal quantum numbers $n = 3-10$, with $T_{eff} = 12000K$ and $\textrm{log}(g) = 8$}
   \label{fig:sestas}
\end{figure}

\begin{figure}
  \includegraphics[width=\textwidth, height=0.8\textwidth]{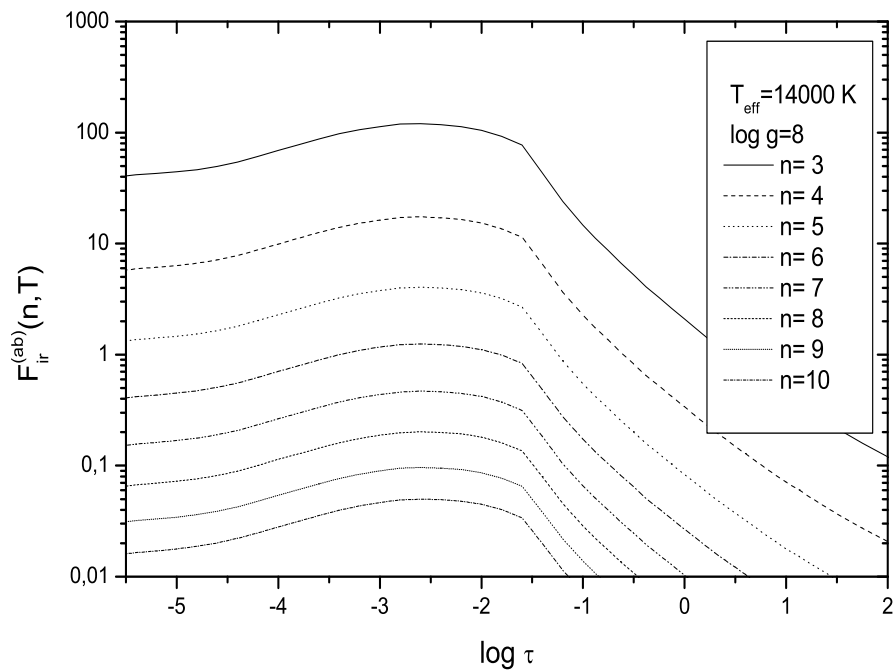}\\
  \caption{Parameter $F_{ir}^{(ab)}(n,T)$ as a function of the logarithm of Rosseland optical
  depth $\textrm{log}(\tau)$, for principal quantum numbers n = 3-10, with $T_{eff} = 14000K$ and $\textrm{log}(g) = 8$}
   \label{fig:sedmas}
\end{figure}

The previous research show that in the helium case the situation
should be similar to the hydrogen case. This assumption is based on
the results obtained in \citet{deveti}, where the processes
(\ref{eq:dva}) and (\ref{eq:cetiri}) were considered from the aspect
of their efficiency in comparison with the other relevant
ionization/recombination processes, for the $n$ from 3 to 10, in the
atmospheres of some of DB white dwarfs. This results are illustrated
by the figures \ref{fig:sestas} and \ref{fig:sedmas} which shows the
behavior of the ratios of the ionization/recombination fluxes caused
by the processes (\ref{eq:dva}) and (\ref{eq:cetiri}) and the
relevant electron-atom and electron-electron-ion processes.

1.  The results presented in the helium case show that the processes
(\ref{eq:dva}) and (\ref{eq:cetiri})  in the atmospheres of the
considered DB white dwarfs should make the similar effects as the
processes (\ref{eq:jedan}) and (\ref{eq:tri}) in the considered M
red dwarf atmosphere.

2.  The processes (\ref{eq:dva} - \ref{eq:cetiri}) should be
included in the models of stelar atmospheres from the beginning in a
consistent way.

\section{Concerning one potential cause of anomalies in Rydberg atoms spectra}

Recent spectroscopic measurements of white dwarfs IR spectra reveal
a gap in the radiation emitted by Rydberg atoms (RA) having values
of the principal quantum number $n \approx 10$ \citep{k1}. Among
possible reasons of these anomalies a number of publications
consider the following processes: (i) collision induced absorption
(CIA), (ii) relativistic quantum effects of "vacuum polarization"
under the super strong magnetic fields $B \ge 10^{13} \textrm{G}$,
(iii) Stark-effect under the electric fields with intensities $E
 \geq 10^{6} \textrm{V/cm}$.

The threshold of the electric field intensity for an auto ionization
process is equal to $E \approx 3 \cdot 10^{4} \textrm{V/cm}$ for RA
with $n > 10$. This means that the lines in RA spectra emitted with
$n \geq 10$-states should be blocked with a strong electric field.
The RA lifetime $\tau_{CH}$ is formed mainly due to intensive chemi
ionization collisions and appears to be $\leq 10^{-8}\textrm{s}$ $(n
\approx 10)$ if the following realistic conditions are realized in
white dwarfs atmospheric plasma: the concentration of ground state
particles $N_{0} \ge 3 \cdot 10^{17} \textrm{cm}^{-3}$, the electron
concentration $N_{e} \ge 10^{8}\ \textrm{cm}^{-3}$, RA concentration
$N_{R}^{*} \ge 10^{13} \textrm{cm}^{-3}$, and the constant $k$ of
the chemi ionization reaction (CHR) $k \approx
10^{-9}\textrm{cm}^{3}\textrm{s}^{-1}$.

\begin{equation}
\label{eq:k} RA + A \to A_{2}^{+} + e
\end{equation}

The mentioned value of the collision lifetime $\tau_{CH}$ is seen to
be by two orders of magnitude less than the RA's reference radiation
values $\tau_{R}$ occurring due to photon spontaneous emission.
Note, moreover, that maximum $k(n)$ corresponds to $n \approx 10$
(see \citet{k2}), that should bring to "selectivity" of the process
(\ref{eq:k}) by $n$ when an external electric field is absent.
\begin{figure}
  \includegraphics[width=\textwidth, height=0.8\textwidth]{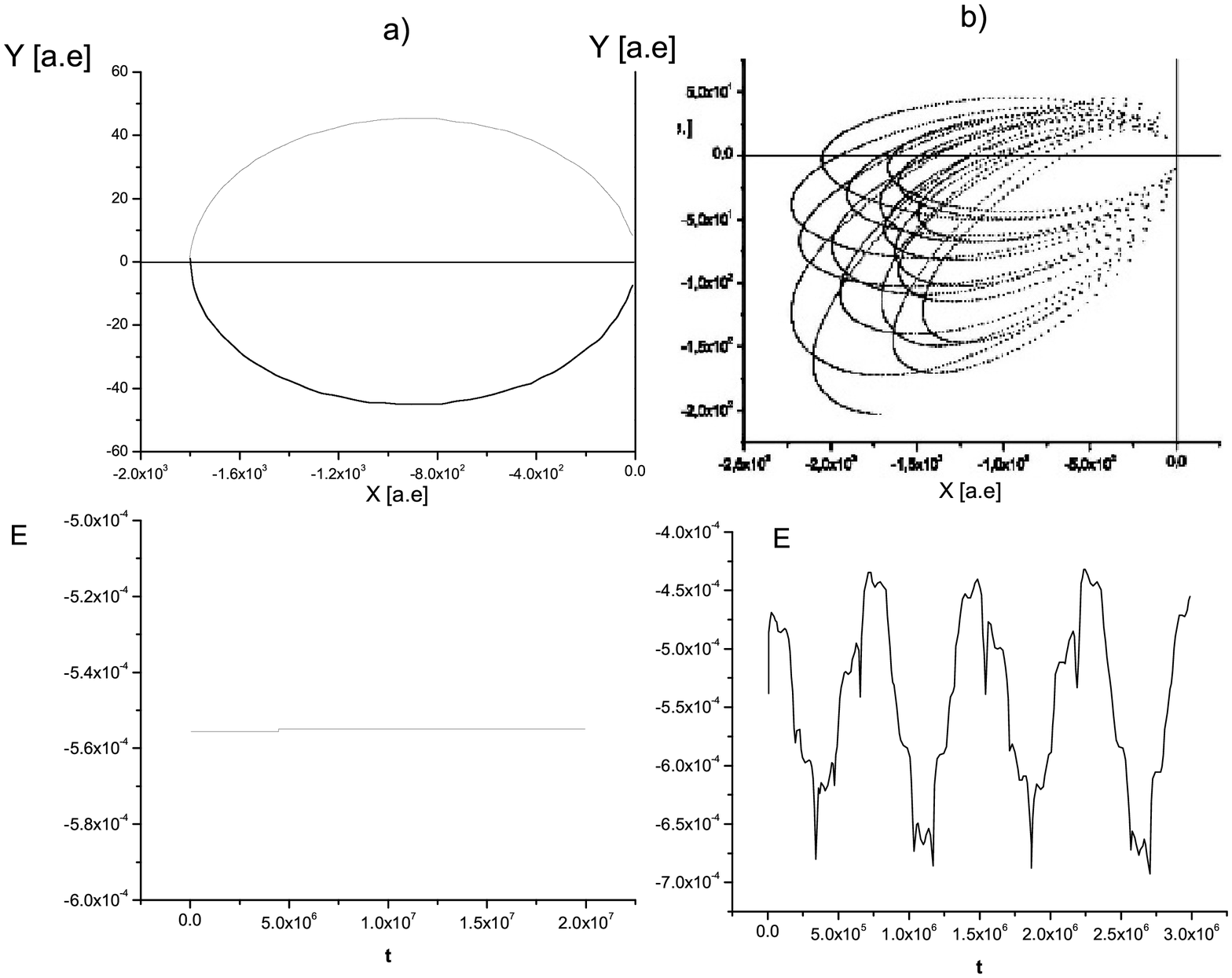}\\
  \caption{RE trajectories (upper Figs) and RE energy dynamics (lower Figs) in the
           hydrogen atom for free atom (a) and in the presence of a microwave field (b).}
   \label{fig:Kfig1}
\end{figure}

\begin{figure}
  \centering
  \includegraphics[width=0.7\textwidth, height=0.6\textwidth]{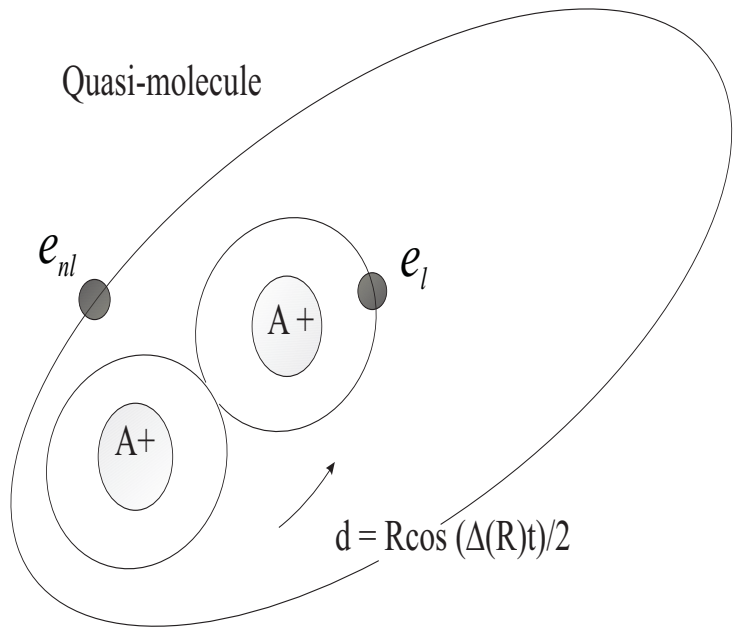}
  \caption{Formation of a microwave electric field in the quasi
           molecular complex due to the exchange of the inner valence electron.  }
   \label{fig:Kfig2}
\end{figure}

\begin{figure}
  \includegraphics[width=\textwidth, height=0.8\textwidth]{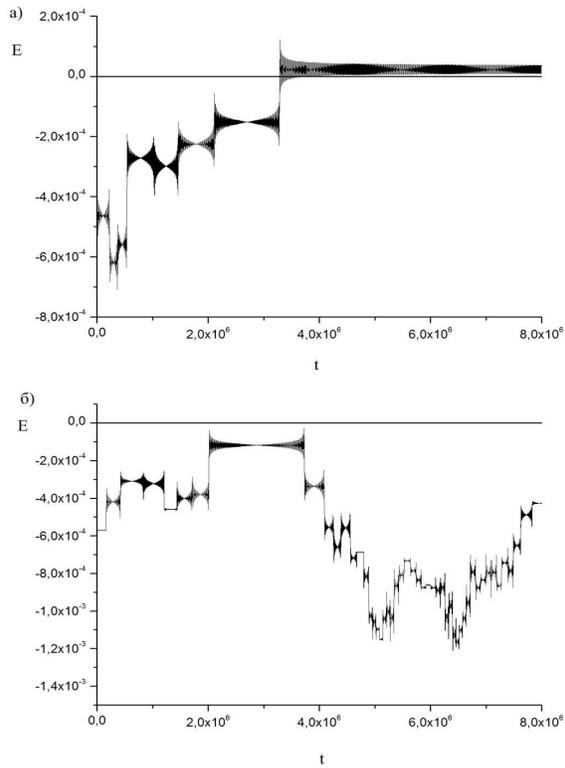}
  \caption{(a) Random dynamics of RE energy E in the hydrogen atom due to coupling with an
           external microwave field resulting in the diffusion (chaotic) ionization $(E>0)$.
           (b) The same for the model Zommerfield atom under the realization of F\"{o}ster
           (the double Stark) resonance. The details of the interatomic potential (\ref{eq:U}) are
           seen to dramatically modify the chaotic dynamics (absence of ionization).}
   \label{fig:Kfig3}
\end{figure}

We'd like to pay attention to one more possibility of RA
reallocation within excited states resulted from the Rydberg
electron (RE) motion chaotization in a single $RA+A$ collision
event. The nature of the so called stochastic dynamics arises due to
nonlinear energy behavior of RE as a function of its principal
quantum number $n$. In an external microwave field the most
important contribution to the coupling between the atom and the
field occurs in the vicinity of the atomic core where the optical
electron has the maximum velocity. Even small energy changes result
in strong motion period variations for RE. The latter leads to a
comparatively rapid (in the scale of the period) dephasing within
the electron and the microwave field oscillations and to the RE
dynamic randomization (Fig. \ref{fig:Kfig1}). As it was shown in an
accurate CHR treatment \citep{k2} the processes (\ref{eq:k}) goes
through the formation of the intermediate quasi molecular complex
$(AA^{+}) + e$,  (see Fig. \ref{fig:Kfig2}), in which RE $e_{nl}$
becomes common for the nuclei. The two atomic ions $A^{+}$ exchange
the inner valence electron $e_{1}$ and produce a variable dipole
moment that oscillates at the exchange interaction frequency. The
latter perturbs RE with an effective microwave electrical field and
eventually manifests itself in the diffusion ionization (Fig.
\ref{fig:Kfig3}a).

The key point of our treatment related to astrophysical applications
is the observation that an external statistic magnetic or electric
field may strongly modify the futures of the dynamic chaos (see Fig.
\ref{fig:Kfig3}) if the atomic state $\{n,l\}$ of RE satisfies the
so called F\"{o}ster or double Stark resonance. The latter is
realized when the RE energy $E_{nl}$ appears to locate exactly
between any two adjoining states belonging to $l'=l \pm 1 $ series:
$2E_{nl} \equiv E_{n'l'}+ E_{n+1'l'}$.

\begin{equation}
\label{eq:U} U_{Z}(r) = - \frac{z}{r} + \frac{a}{2 r^{2}}
\end{equation}

\begin{figure}
  \includegraphics[width=\textwidth, height=0.8\textwidth]{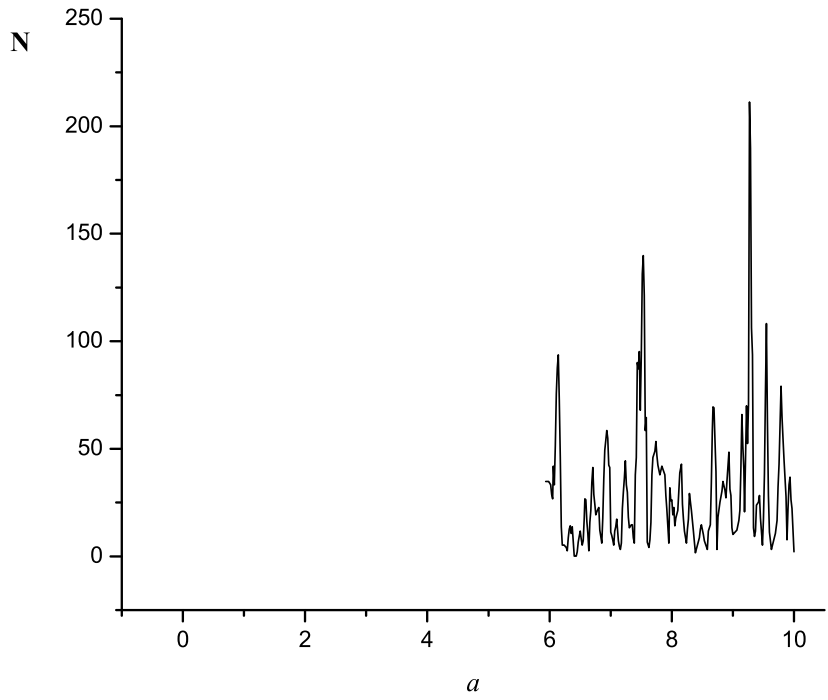}\\
  \caption{Dependence of the diffusion lifetime $(\tau_{CH}) \sim N$ (the number of the RE
           orbital rounds) on the parameter $a$ of the model Zommerfeld atomic potential (\ref{eq:U}).
           The peaks correspond to occurring different F\"{o}ster resonances. The value
           $a=0$ corresponds to the hydrogen atom case.}
   \label{fig:Kfig4}
\end{figure}

Figs. \ref{fig:Kfig3} and  \ref{fig:Kfig4} demonstrate the RE
stochastic dynamics in the model Zommerfield potential $U_{Z}$
(atomic units) which covers the hydrogen case $(z=1, \quad a=0)$.
Zommerfield atom is known to well describe real alkali atoms and it
provides a simple analysis (both analytical and numerical) of the
F\"{o}ster resonance situation. Importantly, in the vicinity of the
F\"{o}ster resonance the diffusion coefficients of the Fokker-Plank
equation describing RE random walk within the energy levels
(Fig.\ref{fig:Kfig2}) appears to be very sensitive to details of the
internuclear potential. One should expect, correspondingly, an
irregular (anomaly) behavior of the diffusion auto ionization time
$\tau_{CH}$ for reaction (\ref{eq:k}) on the intensities of the
external statistic fields. Since the auto ionization time
$\tau_{CH}$ determines the concentration of RAs in white dwarfs
atmospheric plasmas, its anomalies should directly manifest
themselves in the anomalies of RA spectra.

\section{Conclusions}
In this paper it was shown that the atoms and molecules in Rydberg's
states play an are important role in the astrophysics from at least
two reasons:

\begin{itemize}

\item They are large and weakly bound, because the probable astrophysical
targets for searching Rydberg atoms are the cold stars and cosmic
objects including standard Big Bang chemistry, late M-dwarfs, brown
dwarfs and white dwarfs. The basic effect producing by Rydberg atoms
in atmospheres of these stars is significant flux deficits in
Spitzer observations of these stars. In cool stars atom-Rydberg atom
chemi-ionization collision processes become very important. It was
shown also that so-called regime of dynamic chaos should be
considered as typical rather than exceptional situation in Rydbeg
atoms collisions.
\item Much of their atomic structure and behavior in
external fields can be understood on the basis of straightforward
extension of hydrogenic theory. For example, it is interesting to
consider the behavior of Rydberg atoms and molecules in the
atmospheres of the white dwarfs with extremely strong magnetic
fields. In such strong magnetic fields Rydberg atoms become
anisotropic producing special features in the spectra of polarized
radiation.

\end{itemize}

\section{Acknowledgments}

The presented work is part of the project RFBR (No.07-02-00535a) and
Programs of Presidium  and of the Department of Physical Sciences of
RAS, and project 141033 "Non-ideal laboratory and ionosphere plasma
: properties and applications" financed by the Ministry of Science
of the Republic of Serbia, as well as by Russian Foundation for
Basic Research Grant 05-03-33252.

\bibliographystyle{elsarticle-harv}

\end{document}